\documentclass[aip,graphicx]{revtex4-1}
\pdfoutput=1
\usepackage{CJKutf8}
\usepackage{multirow}
\usepackage{amsmath}
\usepackage{cases}
\usepackage{graphicx}
\usepackage{dcolumn}
\usepackage{bm}
\draft 

\usepackage{siunitx}
\usepackage[utf8]{inputenc}
\usepackage[T1]{fontenc}
\usepackage{mathptmx}
\usepackage{etoolbox}
\usepackage{hyperref}

\makeatletter
\def\@email#1#2{%
	\endgroup
	\patchcmd{\titleblock@produce}
	{\frontmatter@RRAPformat}
	{\frontmatter@RRAPformat{\produce@RRAP{*#1\href{mailto:#2}{#2}}}\frontmatter@RRAPformat}
	{}{}
}%
\makeatother

\begin{document}
\raggedbottom

\title{Effect of a horizontal magnetic field on the melting of low Prandtl number liquid metal in a phase-change Rayleigh-Bénard system} 



\author{Xinyi Jiang\begin{CJK}{UTF8}{gbsn}(蒋鑫宜)\end{CJK}}
\author{Chenyu You\begin{CJK}{UTF8}{gbsn}(游晨宇)\end{CJK}}
\author{Xinning Nan\begin{CJK}{UTF8}{gbsn}(南忻宁)\end{CJK}}
\author{Jiawei Chang\begin{CJK}{UTF8}{gbsn}(常家伟)\end{CJK}}
\author{Zenghui Wang\begin{CJK}{UTF8}{gbsn}(王增辉)\end{CJK}}
 \email[]{wzhawk@ucas.ac.cn}
\affiliation{School of Engineering Science, University of Chinese Academy of Sciences, Beijing 100049, China}


\date{\today}

\begin{abstract}
We investigated the low Prandtl number Rayleigh-Bénard (RB) system with a melting top boundary under horizontal magnetic fields. This study is crucial to gain physical insights into the melting dynamics of thermal storage systems, which will help in controlling them. A three-dimensional solid-liquid phase change Rayleigh-Bénard system of gallium was numerically simulated using the enthalpy-porosity method in a cubic domain. In the absence of a magnetic field, the melting process was clearly divided into four regimes: conduction, stable growth, coarsening, and chaotic regimes. We analyzed the flow and heat transfer characteristics in each regime and established scaling relations for the Nusselt number and liquid fraction. Under horizontal magnetic fields, a quasi-two-dimensional (Q2D) flow pattern was observed. We further examined the effects of magnetic field strength on different melting regimes. With increasing Hartmann number, a new flow mode emerged during the stable growth regime. The results also show that the magnetic field alters the relative duration of each melting regime in the overall melting process.
\end{abstract}


\maketitle 

\section{\label{sec:level1}introduction}

With the advancement of modernization, the global demand for energy continues to increase. At present, fossil fuels remain the dominant source of energy. However, fossil fuels are non-renewable and have limited reserves. In addition, their usage leads to environmental pollution and global warming, which threaten human survival. Therefore, the development of new energy sources is one of the key solutions to the energy crisis and environmental challenges.

Compared to other forms of renewable energy, nuclear fusion is less affected by external conditions such as time, space, and climate. It also offers high energy density and reliable performance. Controlled nuclear fusion uses abundant and clean fuel. Deuterium can be extracted from seawater, while tritium can be produced through the reaction of neutrons with lithium. Both the fuel and products are non-radioactive, and the reaction does not emit polluting gases. Hence, controlled nuclear fusion is considered one of the most promising methods to address global energy and environmental issues. The Tokamak is currently the most widely used device for achieving thermonuclear fusion. Its first wall faces the plasma directly and must withstand intense heat loads and high-energy particle bombardment. Common first-wall materials include beryllium, carbon-based materials, and tungsten-based materials. Under extremely high heat flux, these materials may melt or evaporate, which can contaminate the plasma core and increase the operational risks of the reactor.\cite{lipschultz_divertor_2012} Therefore, studying solid-liquid phase change of liquid metals under magnetic fields provides valuable insights for the design of plasma-facing components such as the first wall.

In addition, the development of energy storage technologies is another important approach to addressing the energy crisis and environmental issues. Phase change energy storage using liquid metals has attracted wide attention due to its high energy density and stable operating temperature. In various fields, magnetic fields are often applied to regulate the solid–liquid phase change process of liquid metals. For example, in thermal-related applications such as cooling of compact electronic devices and phase change thermal storage systems, magnetic fields can be adjusted to enhance heat transfer and control phase change duration.\cite{ge_low_2013} Therefore, investigating the flow behavior, heat transfer characteristics, and phase change rate of liquid metals under magnetic fields is of great significance for applications in energy storage, thermal management, metallurgy and biomedicine.\cite{gao_phase_2024,chadha_phase_2022,wang_phase_2022}

Current research on solid-liquid phase change primarily focuses on sidewall heating or cooling configurations, where the imposed temperature gradient is perpendicular to gravity.\cite{mallya_buoyancy-driven_2021} In contrast, studies on phase change in Rayleigh-Bénard systems with bottom heating and top cooling are relatively scarce and mostly concentrate on high-Prandtl-number fluids. Madruga et al.\cite{madruga_experimental_2018} conducted experimental and numerical investigations on the melting of tetradecane inside a cube heated from the bottom. Their study compared experimental and numerical results, with a focus on the evolution of velocity fields, temperature fields, and the solid–liquid interface during the four distinct melting stages: conduction, steady growth, coarsening, and turbulence. Favier et al.\cite{favier_rayleighbenard_2019} numerically simulated a two-dimensional RB melting system with an upper melting boundary using a phase-field method. They identified baroclinic effects caused by the non-planar melting front and quantitatively analyzed the influence of the Stefan number on the interface evolution. Yang et al.\cite{yang_morphology_2023} focused on the morphology of the solid–liquid interface. Their results showed that the interface roughness increases significantly with the liquid layer height, while the Stefan number has a relatively weak influence on it. Ravichandran et al.\cite{ravichandran_combined_2022} further examined the effects of buoyancy, rotation, and shear on the evolution of the solid–liquid phase boundary in a three-dimensional bottom-heated cavity.

However, studies on phase change RB systems involving low-Prandtl-number liquid metals remain limited. Compared with high $Pr$ fluids, low $Pr$ fluids represented by liquid metals have much higher thermal diffusivity than momentum diffusivity, resulting in heat transfer dominated by conduction. In phase-change RB systems, such fluids exhibit a large temporal separation of scales, and their flow and heat transfer characteristics vary significantly across different melting stages. Guo et al.\cite{guo_semi-theoretical_2021} compared the melting processes of phase-change materials with high and low Prandtl numbers. Hasan et al.\cite{hasan_evolution_2021} investigated the effects of Rayleigh number, Stefan number, and aspect ratio on the evolution of the solid–liquid interface and convective heat transfer in low Prandtl number phase-change Rayleigh-Bénard convection within a two-dimensional bottom-heated cavity. Most of the related numerical simulations have been conducted using two-dimensional models, with limited studies focusing on three-dimensional simulations and the detailed analysis of different melting stages. In addition, when an external magnetic field is applied, magnetohydrodynamic (MHD) effects emerge. As electrically conductive fluids, liquid metals experience Lorentz forces while moving in a magnetic field, which can significantly alter their flow and heat transfer behavior. 

Therefore, further investigation is needed to understand both the phase-change dynamics of melting regimes in liquid metal RB systems and the influence of magnetic fields. The main objectives of this study are set to: i) investigate the melting dynamics of low-Prandtl-number liquid metals in a phase change RB system, ii) examine the effects of magnetic fields on the phase change RB system involving liquid metals, and iii) identify the scaling laws for melting and heat transfer.
This paper is organized into four main sections. Section 2 presents the physical model, governing equations, and the verification of mesh independence and model accuracy. Section 3 analyzes the distinct melting stages in the phase-change RB system without a magnetic field, as well as the flow, heat transfer, and melting characteristics of the system under the influence of a magnetic field. Section 4 summarizes the key findings of this study and outlines potential directions for future research.

\section{\label{sec:level1}method}

\subsection{\label{sec:level2}Physical model}

\begin{figure}[htbp]
	\centering
	\includegraphics[width=0.5\textwidth]{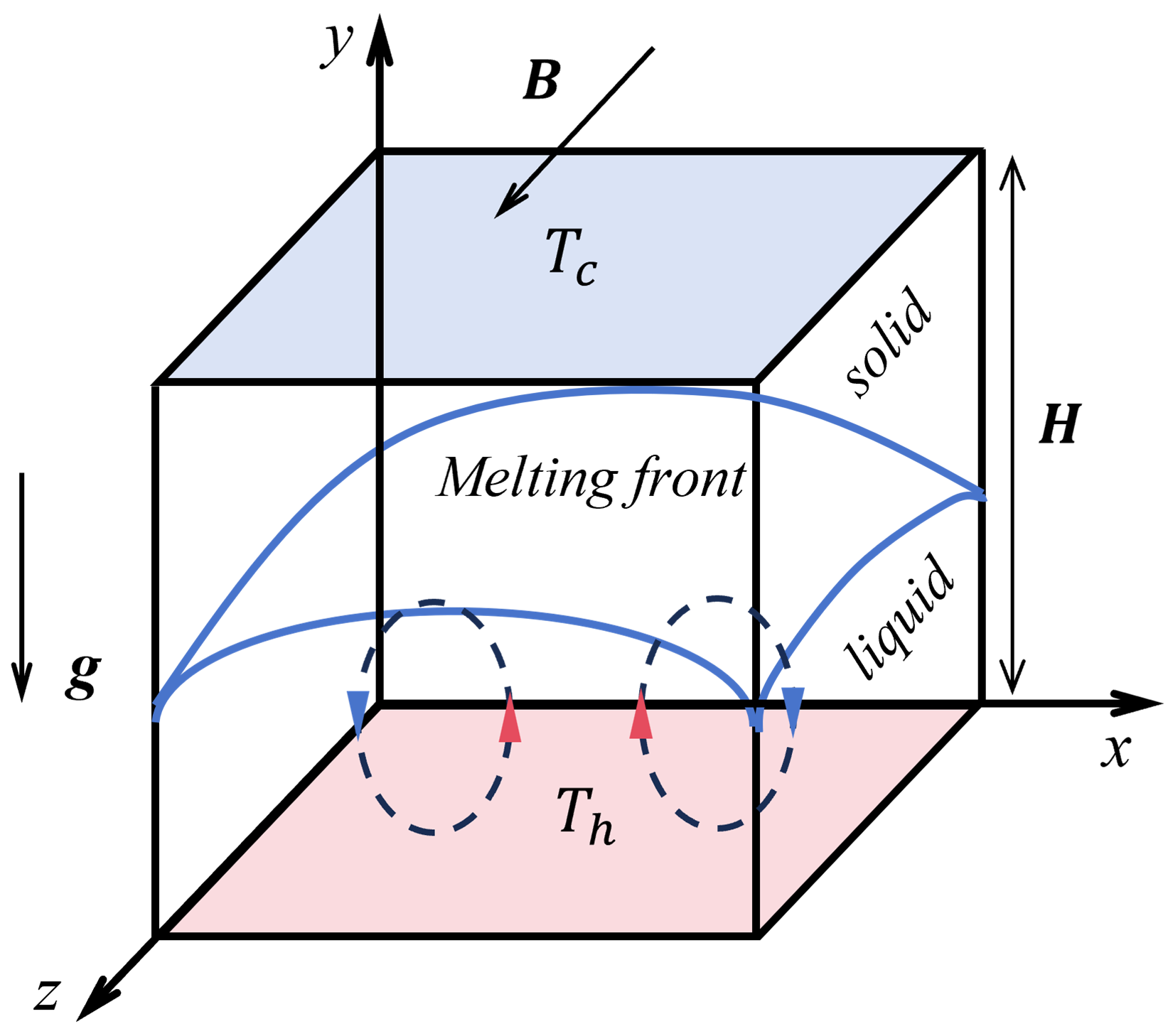}
	\caption{\label{fig:fig1}Schematic diagram of the physical model.}
\end{figure}

This study investigates the natural convection-driven melting of solid gallium in a bottom-heated cubic cavity under a uniform magnetic field. The physical model used in the simulations is illustrated in 
Fig.~\ref{fig:fig1}%
. The boundary conditions are as follows: the bottom and top walls of the cavity are maintained at temperatures $T_h$ and $T_c$, respectively. The initial solid temperature $T_0$ is equal to the top wall temperature $T_c$. The sidewalls are adiabatic, and all walls are assumed to be electrically insulated and no-slip. A uniform horizontal magnetic field is applied in the $z$-direction.

The Boussinesq approximation is employed, where density variation with temperature is considered only in the buoyancy term, and all other thermophysical properties are treated as constants. The density is assumed to be the same in the solid and liquid phases ($\rho_s = \rho_l$). 
Because the magnetic Reynolds number of the liquid metal is very small ($Re_m = \mu \sigma L u \ll 1$), the externally applied magnetic field is much stronger than the induced magnetic field generated by the electric current. Therefore, magnetic induction effects are neglected. The thermophysical properties of gallium used in this study are listed in Table~\ref{tab:table1}.
\begin{table}
	\caption{\label{tab:table1}Physical properties of gallium.  }
	\begin{ruledtabular}
		\begin{tabular}{lcr}
			Property&Unit&Value\\
			\hline
			Density $\rho$ & \si{kg/m^3} & 6093\\
			Specific heat $c_p$ & \si{J/(kg \cdot K)} & 397.6\\
			Thermal expansion coefficient $\beta$ & \si{1/K} & \num{1.27e-4}\\
			Kinematic viscosity $\nu$ & \si{m^2/s} & \num{3.15e-7}\\
			Thermal diffusivity $\alpha$ & \si{m^2/s} & \num{1.29e-5}\\
			Electrical conductivity (solid) $\sigma_s$ & \si{S/m} & \num{6.64e6}\\
			Electrical conductivity (liquid) $\sigma_l$ & \si{S/m} & \num{3.85e6}\\
			Latent heat of fusion $L$ & \si{J/kg} & 80160\\
		\end{tabular}
	\end{ruledtabular}
\end{table}

\subsection{\label{sec:level2}Governing Equation}

The governing equations are nondimensionalized using the following characteristic scales: cavity height $H$ for length, $\rho H^3$ for mass, $\alpha /H$ for velocity, $H^2/\alpha$ for time, $B$ for magnetic field strength, and $\Delta T = (T_h-T_c)$ for temperature. The resulting dimensionless governing equations are as follows:
\begin{eqnarray}
	\nabla\cdot\bm{u} = 0
	\label{eq:one},
	\\
	\frac{\partial \bm{u}}{\partial t}+\nabla \cdot (\bm{uu}) = -\nabla p + RaPr\theta \bm{e}_y + Pr\nabla^2\bm{u} - A\bm{u} + Ha^2Pr(\bm{j} \times \bm{e}_B)
	\label{eq:two},
	\\
	\bm{j} = -\nabla \varphi + \bm{u}\times \bm{e}_B
	\label{eq:three},
	\\
	\nabla \cdot \bm{j} = 0
	\label{eq:four},
	\\
	\frac{\partial \theta}{\partial t}+\nabla \cdot (\bm{u}\theta) = \nabla^2\theta - Ste\frac{\partial f}{\partial t}
	\label{eq:five}.
\end{eqnarray}
where $\bm{u}$, $p$, $\theta = (T-T_c )/(T_h-T_c )$, $\bm{j}$, $\varphi$, and $f$ represent the velocity vector, pressure, temperature, current density, electric potential, and liquid volume fraction from a reference value, respectively. The unit vectors $\bm{e}_y$ and $\bm{e}_B$ denote the directions of gravity and the applied magnetic field, respectively.

The relationship between the liquid fraction $f$ and the temperature $\theta$ is given by:
\begin{align}
	f(\theta)=
	\begin{cases}
		1  &\theta > \frac{2\eta}{\Delta T} \\
		\frac{\theta\Delta T}{2\eta}  &0 \le \theta \le \frac{2\eta}{\Delta T} \\
		0  &\theta \le 0
	\end{cases}
	\label{eq:six}
\end{align}
where $\eta = 0.01\text{K}$.  And $f(\theta_m) = 0.5$ when the temperature is equal to the melting temperature.

The enthalpy-porosity method based on a fixed grid is employed in this study.\cite{voller_fixed_1987} In the porosity method, the cells undergoing phase change are treated as porous media. A Darcy source term $A\bm{u}=C\bm{u} [(1-f)^2⁄(f^3+\epsilon)]$ is added to the momentum equation to suppress flow in the solid phase, based on the Carman-Kozeny relation derived from Darcy’s law.\cite{carman_fluid_1997} The Darcy coefficient is set to $C=\num{1e6}$, and $\epsilon=\num{1e-3}$ is a small stabilizing constant. The enthalpy method treats the latent heat as a source term in the energy equation, which reduces the degree of nonlinearity in the system. To address the inconsistency between the liquid fraction and the temperature in the energy equation, an iterative correction method is applied, as given below:
\begin{eqnarray}
	\frac{\partial \theta^{k+1}}{\partial t}+\nabla \cdot (\bm{u}\theta^{k+1}) = \nabla^2\theta^{k+1} - Ste\frac{\partial f^k}{\partial t}
	\label{eq:seven},
	\\
	f^{k+1} = f^k + \omega \frac{c_p}{L}(\theta^{k+1} - f^{-1} (f^k))
	\label{eq:eight},
	\\
	f^{k+1} = min[max[f^{k+1} , 0], 1]
	\label{eq:nine}.
\end{eqnarray}

First, the temperature of next iteration $\theta^{k+1}$ is computed using the current step liquid fraction $f^k$. Then, the liquid fraction is updated based on the enthalpy formulation. A relaxation factor $\omega$ is applied, and the updated liquid fraction is constrained within the range [0,1].

Using the cavity height $H$ as the characteristic length, the dimensionless Nusselt number $Nu$, Rayleigh number $Ra$, Prandtl number $Pr$, Stefan number $Ste$, Hartmann number $Ha$, mean kinetic energy density $K$, and Fourier number $Fo$ are defined as follows. 
\begin{eqnarray}
	Nu=\iint_{0}^{1}\left.\frac{d\theta}{dy}\right|_{y=0}\mathrm{d}x\,\mathrm{d}z , \quad Ra=\frac{g\beta \Delta T H^3}{\nu\alpha}, \quad Pr = \frac{\nu}{\alpha}, \quad Ste = \frac{c_p \Delta T}{L}
	\nonumber\\
	Ha = BH\sqrt{\frac{\sigma}{\rho\nu}}, \quad K = U_{mag}^{2}, \quad Fo = \frac{\alpha t}{H^2}
	\label{eq:ten}.
\end{eqnarray}
where $B$ is the magnetic field strength and $U_{mag}$ is the dimensionless velocity scalar.

We define the dimensionless solid–liquid interface height $h(x,z,t)$ as the location where the liquid fraction equals 0.5, i.e., $f(x,y=h,z,t)=0.5$.
To better describe the average molten height, we introduce the dimensionless fluid-averaged height: $\bar{h}(t) = \iint_{0}^{1} h(x, z, t) \mathrm{d}x\,\mathrm{d}z = f$, where $f$ is the overall liquid fraction.

To more accurately analyze the system’s melting behavior at different stages, we take $\bar{h}(t)$ as a time-dependent characteristic length and define the effective Nusselt number $Nu_e$ and effective Rayleigh number $Ra_e$ as follows:
\begin{eqnarray}
	Nu_e = \left(\iint_{0}^{1}\left.\frac{d\theta}{dy}\right|_{y=0}\mathrm{d}x\,\mathrm{d}z\right)\bar{h}(t) = Nuf
	\label{eq:eleven},
	\\
	Ra_e = \frac{g\beta \Delta T H^3}{\nu\alpha}\bar{h}(t)^3 = Raf^3
	\label{eq:twelve}.
\end{eqnarray}

\subsection{\label{sec:level2}Numerical Methods}

The numerical solver is developed based on the open-source platform OpenFOAM, using the finite volume method on a non-uniform orthogonal collocated grid to discretize the governing equations.

A customized phase-change solver for liquid metals, with and without magnetic fields, is implemented by modifying the OpenFOAM solver buoyantBoussinesqPimpleFoam. The PIMPLE algorithm is employed to solve the discretized equations. The PISO (Pressure-Implicit Split Operator) method is used for pressure–velocity coupling, and the Rhie-Chow interpolation is applied to eliminate pressure checkerboarding.

To address the inconsistency between the calculated temperature and the temperature–liquid fraction relation in the energy equation, an iterative correction method is employed. The electric current density and Lorentz force are calculated using a compatible conservative scheme.

For temporal discretization, a second-order implicit backward scheme is used. The convective term is discretized using the Gauss linear upwind scheme, and the diffusion term is treated with the Gauss linear corrected scheme.

\subsection{\label{sec:level2}Grid and Numerical Model Validation}

\begin{figure}[htbp]
	\centering
	\includegraphics[width=0.5\textwidth]{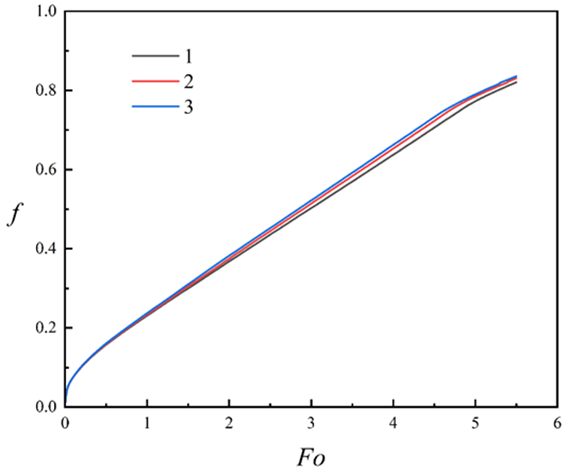}
	\caption{\label{fig:fig2}Variation of liquid phase volume fraction with time under different mesh sizes.}
\end{figure}

A grid independence test was first conducted for the melting of liquid gallium in a 3-D cavity without a magnetic field, under the conditions of $Ra=\num{2e5}$, $Ste=0.0516$, $Pr=0.02249$. Details of the grid configurations are listed in Table~\ref{tab:table2}.
\begin{table}
	\caption{\label{tab:table2}Grid Configurations.}
	\begin{ruledtabular}
		\begin{tabular}{ccccc}
			&$N_x$&$N_y$&$N_z$&Time(min)\footnote{Time: Computational time required for liquid phase volume fraction to reach 0.9}\\
			\hline
			1&90 & 60 & 60 & 393.83 \\
			2 & 120 & 80 & 80 & 2082.32 \\
			3 & 150 & 100 & 100 & 6543.87 \\
		\end{tabular}
	\end{ruledtabular}
\end{table}
, and the validation results are shown in Fig.~\ref{fig:fig2}%
. The average relative error between Grid 1 and Grid 3 is $2.9\%$, and between Grid 2 and Grid 3 is $1.4\%$. Considering both computational accuracy and efficiency, Grid 2 is selected for the simulations. In addition, to account for the presence of Hartmann layers, the mesh is refined near the boundaries to ensure that at least five grid points are included within the Hartmann layer thickness ($\delta_{Ha}=H/Ha$) for all Hartmann numbers studied.

\begin{figure}[htbp]
	\centering
	\includegraphics[width=0.5\textwidth]{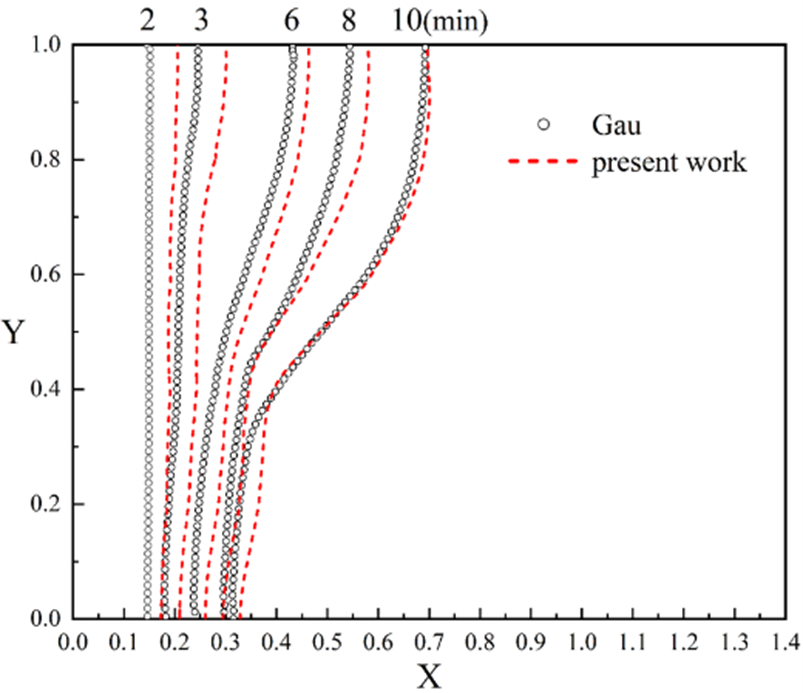}
	\caption{\label{fig:fig3}Comparison of the shape and position of the solid-liquid interface during three-dimensional melting without magnetic field.}
\end{figure}

Additionally, the accuracy of the solid-liquid phase change model was verified. A simulation of gallium melting in a three-dimensional cavity heated from the sidewall was conducted and compared with the experiments of Gau.\cite{gau_melting_1986} The results are shown in Fig.~\ref{fig:fig3}%
, where the solid-liquid interface matches well. Larger discrepancies were observed at the initial melting stage, which is speculated to be caused by experimental measurement errors and the supercooling phenomenon of solid gallium. In high-Prandtl-number systems, due to the low thermal conductivity of the solid phase limiting heat transfer into the solid, the initial solid supercooling has minimal impact. However, for phase change materials with high thermal conductivity, slight initial supercooling can lead to a sharp decrease in melting rate.\cite{webb_analysis_1986}

Then, the natural convection model of liquid metal under magnetic field was validated. The natural convection of molten silicon ($Pr = 0.054$) in a side-heated cavity under magnetic field was simulated, and the obtained average Nusselt number at the heated wall was compared with simulation results from other studies. The comparison results are shown in Table~\ref{tab:table3}.
\begin{table}
	\caption{\label{tab:table3}Comparison of average Nusselt number at the heated wall.}
	\begin{ruledtabular}
		\begin{tabular}{cccccc}
			Ra & Ha & Ozoe\cite{ozoe_effect_1989} & Mo\ss ner\cite{moiner_numerical_1999} & Chen\cite{chen_study_2018} & Present \\
			\hline
			\multirow{3}{*}{$10^6$} & 100 & 4.458 & 4.766 & 4.547 & 4.521 \\
			& 200 & 2.917 & 2.989 & 2.891 & 2.842 \\
			& 300 & 2.251 & 2.245 & 2.175 & 2.100 \\
		\end{tabular}
	\end{ruledtabular}
\end{table}
. The good agreement demonstrates the accuracy of the MHD natural convection model.

\section{\label{sec:level1}results and discussion}

\subsection{\label{sec:level2}Analysis of the Solid-Liquid Phase Change Rayleigh-Bénard System without Magnetic Field}

\begin{figure}[htbp]
	\centering
	\includegraphics[width=0.5\textwidth]{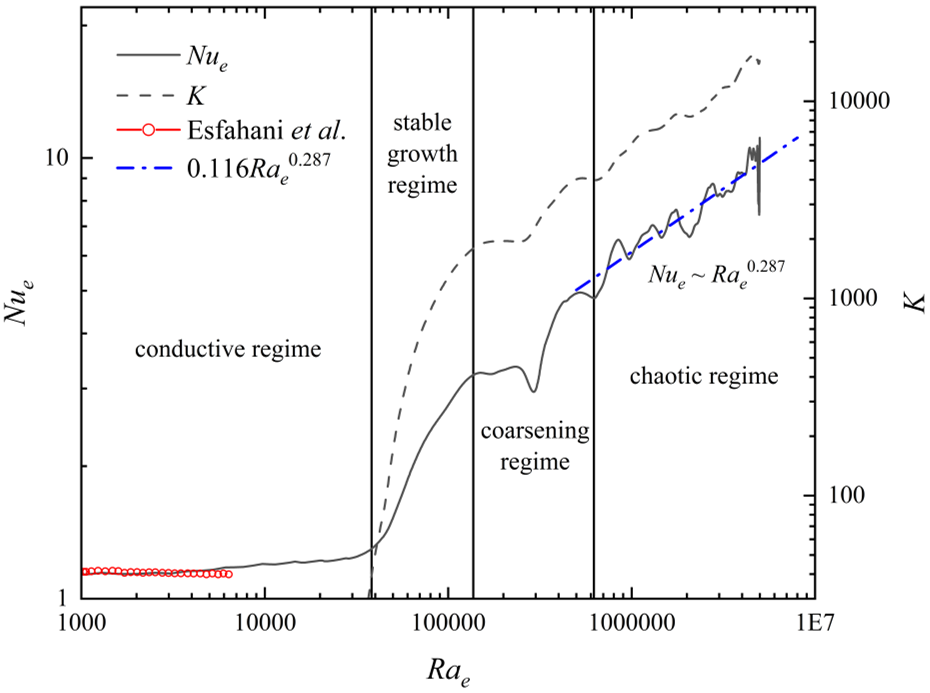}
	\caption{\label{fig:fig4}Variation of the effective Nusselt number Nue at the heated wall (left axis) and the mean kinetic energy density $K$ (right axis) with the effective Rayleigh number at $Ha = 0$. The red line with circular empty symbols corresponds to the effective Nusselt number in the conductive regime reported by Esfahani et al.\cite{esfahani_basal_2018}. The result of linear fitting during the chaotic regime is represented with the blue dashed-dotted line.}
\end{figure}

Previous studies have shown that the phase change Rayleigh-Bénard (RB) system can be divided into four regimes.\cite{favier_rayleighbenard_2019,madruga_dynamic_2018,kansara_probing_2024} Based on the variation of the effective Nusselt number with the effective Rayleigh number, the melting process in this study is divided into four regimes: conductive regime, stable growth regime, coarsening regime, and chaotic regime(Fig.~\ref{fig:fig4}
). For each regime, we analyze the flow structure, heat transfer, and melting rate of the RB system in the absence of a magnetic field. The parameters used are $Ha = 0$, $Ste = 1.264$, $Pr = 0.024$, and $Ra = \num{5e6}$. 

\subsubsection{\label{sec:level3}Conductive Regime}

\begin{figure}[htbp]
	\centering
	\includegraphics[width=0.8\textwidth]{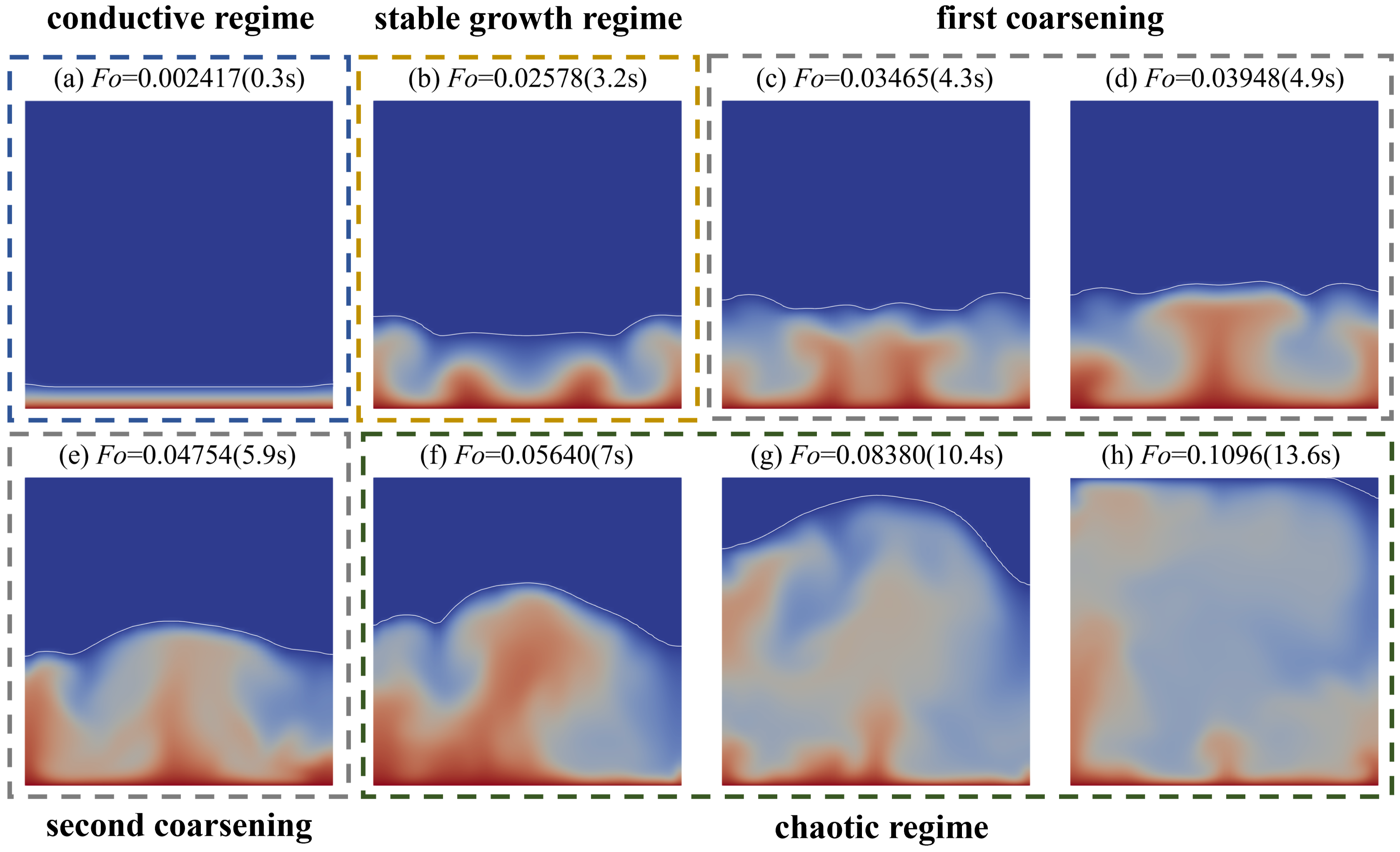}
	\caption{\label{fig:fig5}Temperature fields on the $x\text{-}y$ plane ($z = 0.5$) at $Ha = 0$ for different dimensionless times (Dark red represents $\theta = 1$, and dark blue represents $\theta = 0$). The white solid line corresponds to the solid-liquid interface.}
\end{figure}

\begin{figure}[htbp]
	\centering
	\includegraphics[width=0.5\textwidth]{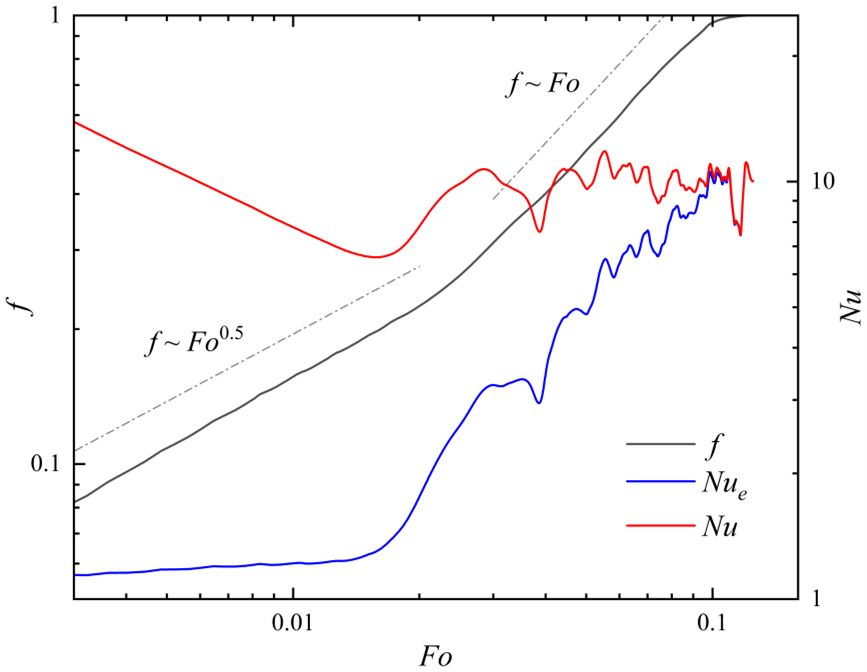}
	\caption{\label{fig:fig6}Variation of the liquid volume fraction $f$ (left axis), Nusselt number $Nu$ and the effective Nusselt number $Nu_e$ at the heated wall (right axis) with dimensionless time $Fo$ at $Ha = 0$. The gray line represents the scaling law between the liquid volume fraction and dimensionless time.}
\end{figure}

The first regime is the conductive regime. As shown in Fig.~\ref{fig:fig5}
(a), there is no noticeable flow inside the melt. The solid-liquid interface remains flat, and the temperature distribution is linear. The isotherms are parallel to the bottom wall. The thermal boundary layer has a thickness of $\delta_T=h/2$, which increases as the interface rises. During this stage, the melting process corresponds to the two-phase Stefan problem on a semi-infinite slab,\cite{alexiades_mathematical_2018} and the analytical solution for the liquid volume fraction is:
\begin{eqnarray}
	f = 2\lambda \sqrt{Fo}
	\label{eq:thirteen}
\end{eqnarray}
as shown in Fig.~\ref{fig:fig6}
, the liquid volume fraction follows the scaling law $f\sim\sqrt{Fo}$. The effective Nusselt number remains constant, representing the conductive regime in Fig.~\ref{fig:fig4}
. The red line with circular open symbols denotes the effective Nusselt number obtained by Esfahani et al.\cite{esfahani_basal_2018} during the conductive regime, one has:
\begin{eqnarray}
	Nu_e = \frac{2\lambda}{\sqrt{\pi}\text{erf}(\lambda)}
	\label{eq:forteen}
\end{eqnarray}
where $\lambda$ is determined by the Stefan number according to
\begin{eqnarray}
	Ste = \sqrt{\pi}\lambda e^{\lambda^2}\text{erf}(\lambda)
	\label{eq:fifteen}
\end{eqnarray}
At the end of this regime, linear Rayleigh-Bénard instability causes the melt to become unstable. Weak convection emerges, and the formation of convection cells can be observed.\cite{madruga_dynamic_2018}

\subsubsection{\label{sec:level3}Stable Growth Regime}

As the solid-liquid interface advances, the interaction between the fluid and the interface becomes stronger. The effective Rayleigh number Rae gradually increases. When it reaches a critical value $Ra_c$, buoyancy becomes strong enough to overcome viscosity and thermal diffusion. The dominant heat transfer mode transitions from conduction to convection, and the flow enters the stable growth regime.
As shown in Fig.~\ref{fig:fig5}(b), four uniformly distributed thermal plumes form above the bottom wall. Each plume corresponds to a pair of counter-rotating convection cells. The number of plumes is determined by the critical wavenumber of the first linear Rayleigh-Bénard instability. This number remains unchanged throughout the stable growth regime.
With continued melting, plume growth becomes apparent. The plumes near the sidewalls grow faster, with elongated stems and widened heads. The central plumes grow more slowly, and their heads bend outward. As the plumes rise and reach the solid-liquid interface, they accelerate local melting. The initially flat interface evolves into a periodic wavy topography. This non-planar topography exerts a feedback on the flow. At any Rayleigh number, it drives baroclinic flow, known as the baroclinic effect.\cite{kelly_thermal-convection_1978}
As convection develops, the thermal boundary layer is disrupted. Its thickness decreases gradually, and the melting rate increases accordingly.

The beginning of the stable growth regime corresponds to a sharp rise in both the effective Nusselt number and the kinetic energy density, as shown in Fig.~\ref{fig:fig4}. At this point, the effective Rayleigh number reaches the critical value. Studies have shown that this critical Rayleigh number depends on $Ste$ and $Pr$. The Stefan number determines the timescale separation between fluid flow and interface motion. A larger $Ste$ implies a larger timescale separation between convection and phase change.\cite{favier_rayleighbenard_2019} In contrast, a larger $Pr$ means that momentum diffuses faster than heat, leading to a smaller timescale separation.\cite{lu_rayleigh-benard_2021}
In the context of phase change RB systems, Vasil et al.\cite{vasil_dynamic_2011} conducted a nonlinear asymptotic analysis and found that the critical Rayleigh number approaches 1295.78 as $Ste$ → 0. Stability analysis by Kim et al.\cite{kim_onset_2008} and numerical simulations by Esfahani et al.\cite{esfahani_basal_2018} also indicate that a larger Ste delays the onset of convection and results in a higher critical Rayleigh number.
In this study, the parameters are $Ste = 1.26$ and $Pr = 0.024$. As a result, buoyancy-driven instability develops slowly, while the melting rate remains relatively high. The timescale separation between convection and interface propagation is large. The observed critical Rayleigh number is approximately $Ra_c\sim10^4$, indicating a delayed onset of convection compared to classical RB systems. This convective delay has been reported in many previous studies.\cite{favier_rayleighbenard_2019,kansara_probing_2024,esfahani_basal_2018,lu_rayleigh-benard_2021,li_convection_2023}

\subsubsection{\label{sec:level3}Coarsening Regime}

For cavities with a size smaller than 0.05 m,\cite{madruga_dynamic_2018} as melting progresses, convection cells continue to grow. The horizontal flow of the fluid is restricted by the sidewalls, which induces horizontal shear flow. This shear disturbs the primary convection structure and triggers secondary Rayleigh-Bénard instability. The system then enters the coarsening regime.
As shown in Fig.~\ref{fig:fig5}(c)(d), during this regime, the central thermal plumes begin to expand, while the plumes near the sidewalls converge. Eventually, the two central plumes merge into one, forming a new convection cell. The aspect ratio of the convection cell remains nearly constant throughout this process. During coarsening, the periodic structure of the solid-liquid interface is disrupted, and the amplitude of interface fluctuations decreases. As shown in Fig.~\ref{fig:fig4}, $Nu_e$ and $K$ fluctuate irregularly in this regime, without a clear trend.
After the first coarsening event, the system does not transition directly into a chaotic state. Instead, a second coarsening event is observed, which can be considered a coarsening cascade. As shown in Fig.~\ref{fig:fig5}(e), the central and left-side plumes continue to expand and rise, while the right-side plume gradually shrinks. 
Whether a second coarsening event occurs depends on the size of the cavity. In large cavities, there is more space for interactions between phase separation and convective motion before full melting is reached. Therefore, the system tends to enter a chaotic regime without undergoing a second coarsening. For example, in the study by Madruga et al.\cite{madruga_dynamic_2018}, a cavity with a size of 0.04 m exhibited a second coarsening event, which is consistent with the observations in the present study.

\subsubsection{\label{sec:level3}Chaotic Regime}

After the coarsening regime, thermal plumes begin to shift noticeably and gradually dissipate. As time progresses, the convective region continues to develop, and the flow field exhibits strong temporal fluctuations. Irregular and fine thermal plumes are generated from the heated bottom wall. These new plumes are independent of the previous convection cells.
The thermal boundary layer first thins rapidly, then undergoes oscillations. Due to the highly unsteady flow field, the solid-liquid interface experiences significant deformation.
A scaling analysis of this regime shows that the liquid volume fraction follows the relation $f \sim Fo$, as illustrated in Fig.~\ref{fig:fig6}. From Fig.~\ref{fig:fig4}, it can be observed that the effective Nusselt number exhibits continuous oscillations. A linear fit yields the correlation shown by the blue line: $Nu_e=0.116Ra_{e}^{0.287}$.

\subsection{\label{sec:level2}Analysis of Phase Change RB System under Magnetic Field}

This section focuses on analyzing the effects of magnetic field strength on the flow, heat transfer, and melting behavior of the phase change RB system. The parameters used in this section are $Ha$ = 0, 25, 50, and 100; $Ste = 1.264$; $Pr = 0.024$; and $Ra = \num{5e6}$.

\subsubsection{\label{sec:level3}Quasi-Two-Dimensionalization of Flow under Magnetic Field}

\begin{figure}[]
	\centering
	\includegraphics[width=0.7\textwidth]{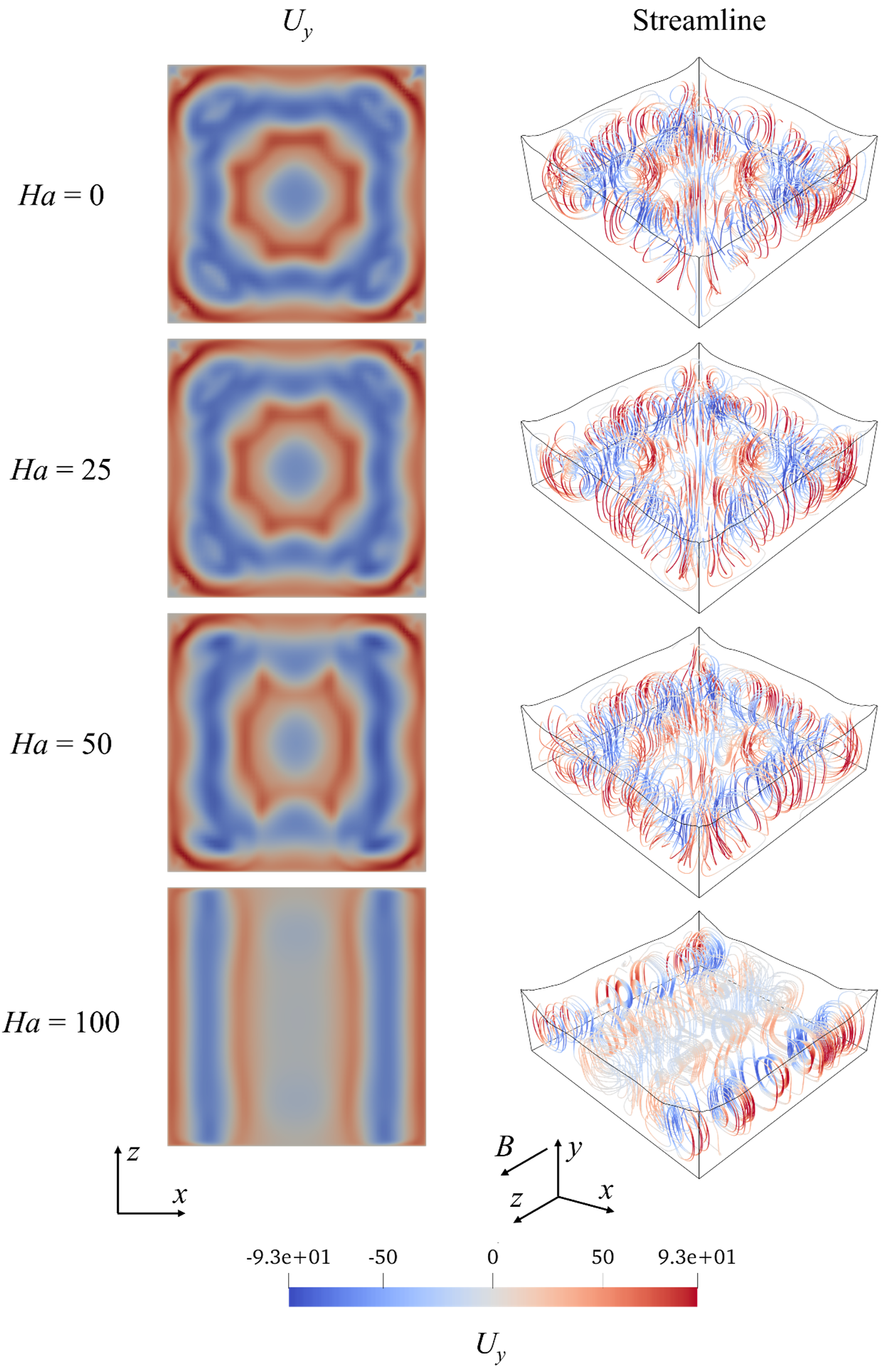}
	\caption{\label{fig:fig7}Vertical velocity fields and streamlines on the $x\text{-}z$ plane ($y=0.05$) at $Fo=0.02578$ under different $Ha$. The outer black solid line indicates the interface between the molten region and the cavity. The color map represents the vertical velocity component $U_y$, with red and blue corresponding to upward and downward flow respectively.}
\end{figure}

As shown in the streamline and vertical velocity field plots at $Fo=0.02578$(Fig.~\ref{fig:fig7}
), the flow without magnetic field exhibits a fully three-dimensional structure, which is referred to as cell structures by Vogt et al.\cite{vogt_free-fall_2021}. The convection cells near the cavity diagonals are inclined toward the center, while those near the mid-plane flow parallel to the sidewalls.
Under a horizontally uniform magnetic field applied along the z-direction, flow components aligned with the magnetic field are suppressed. Consequently, the flow structure transitions from a 3-D convective cell pattern to quasi-2-D convective roll pattern, consistent with the observations in Rayleigh-Bénard systems without phase change under magnetic fields.\cite{fauve_effect_1981,chen_effects_2024,burr_rayleighbenard_2002,yanagisawa_convection_2013,vogt_transition_2018,yang_transition_2021,wu_influence_2025}

At $Ha=25$, the influence of the magnetic field remains limited. The velocity and streamline plots show that buoyancy still dominates, and the flow retains a 3-D convective cell pattern circulating toward the center.
At $Ha=50$, the suppression effect of the magnetic field becomes more pronounced. The vertical velocity field on the $x\text{-}z$ plane shows that the flow near the mid-plane ($x=0.5$) is mainly oriented in the $y\text{-}z$ direction. Convection along the $z$-direction weakens, and the inclined convective cells along the center diagonal begin to spread along the magnetic field direction. The flow starts to reorganize into convective rolls aligned within the $x\text{-}y$ plane. Convection along the $x$-direction is enhanced, as indicated by increased vertical velocity near the sidewalls, suggesting that the magnetic field acts as a flow rectifier.
At $Ha=100$, the RB convection becomes fully quasi-two-dimensional, forming convection rolls parallel to the $x\text{-}y$ plane. The vertical velocity field exhibits a striped distribution, with convective rolls extending along the magnetic field direction. The velocity of the central convective rolls is lower than that of the rolls near the sidewalls. This is because the central flow near $x=0.5$ is mainly aligned with the $y\text{-}z$ plane and is thus more strongly suppressed by the magnetic field. As a result, the vertical velocity near the centerline is much lower than that near the sidewalls. In addition, with increasing Hartmann number, the Lorentz force becomes dominant, leading to a further reduction in the vertical velocity of the wall-adjacent rolls.

\begin{figure}[]
	\centering
	\includegraphics[width=0.7\textwidth]{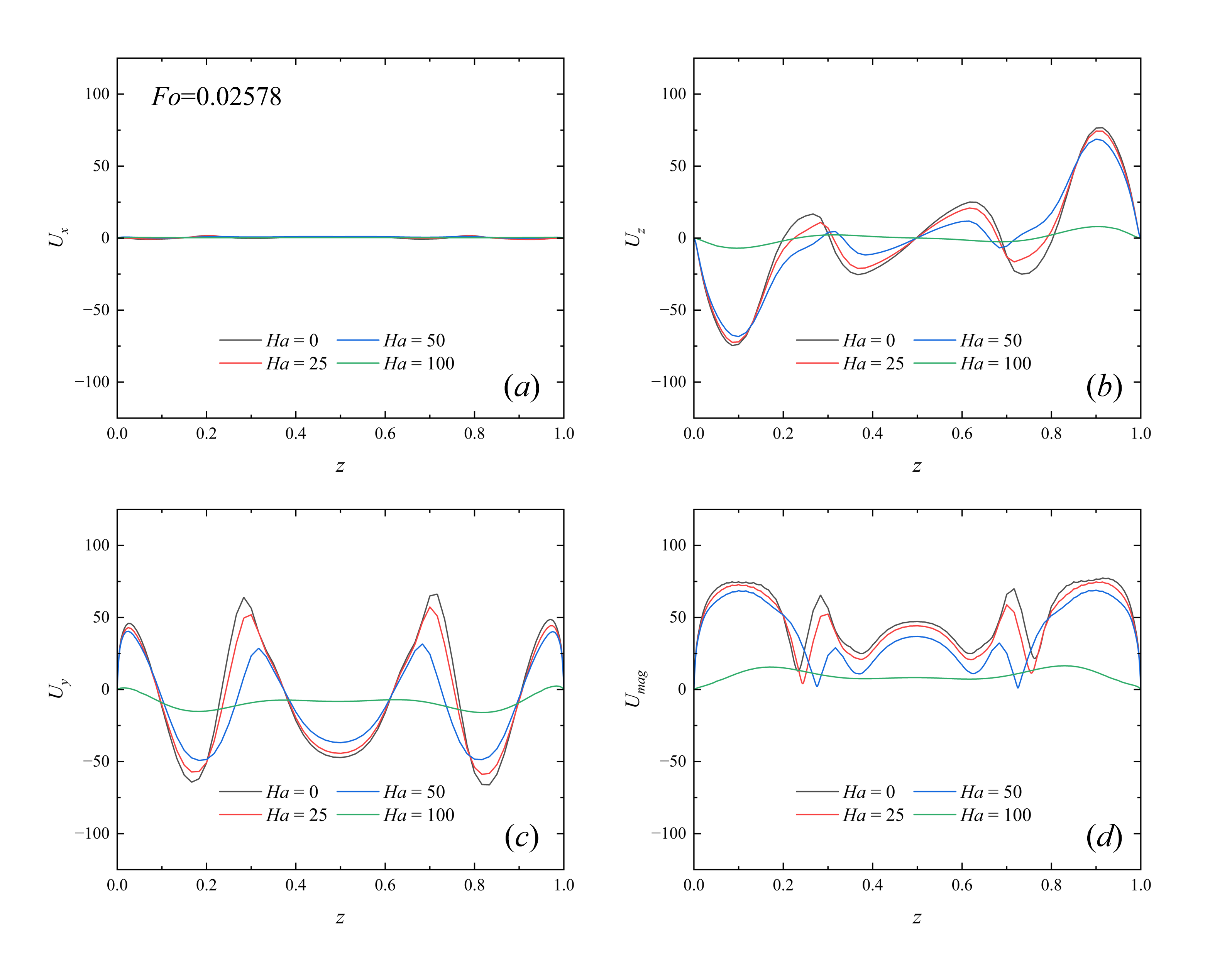}
	\caption{\label{fig:fig8}Distributions of velocity components at the $x$-axis midline near the bottom heated wall ($y=0.05$, $x=0.5$) at $Fo=0.02578$ under different Hartmann numbers: (a) $U_x$ (b) $U_z$ (c) $U_y$ (d) $U_{mag}$.}
\end{figure}

\begin{figure}[]
 	\centering
 	\includegraphics[width=0.7\textwidth]{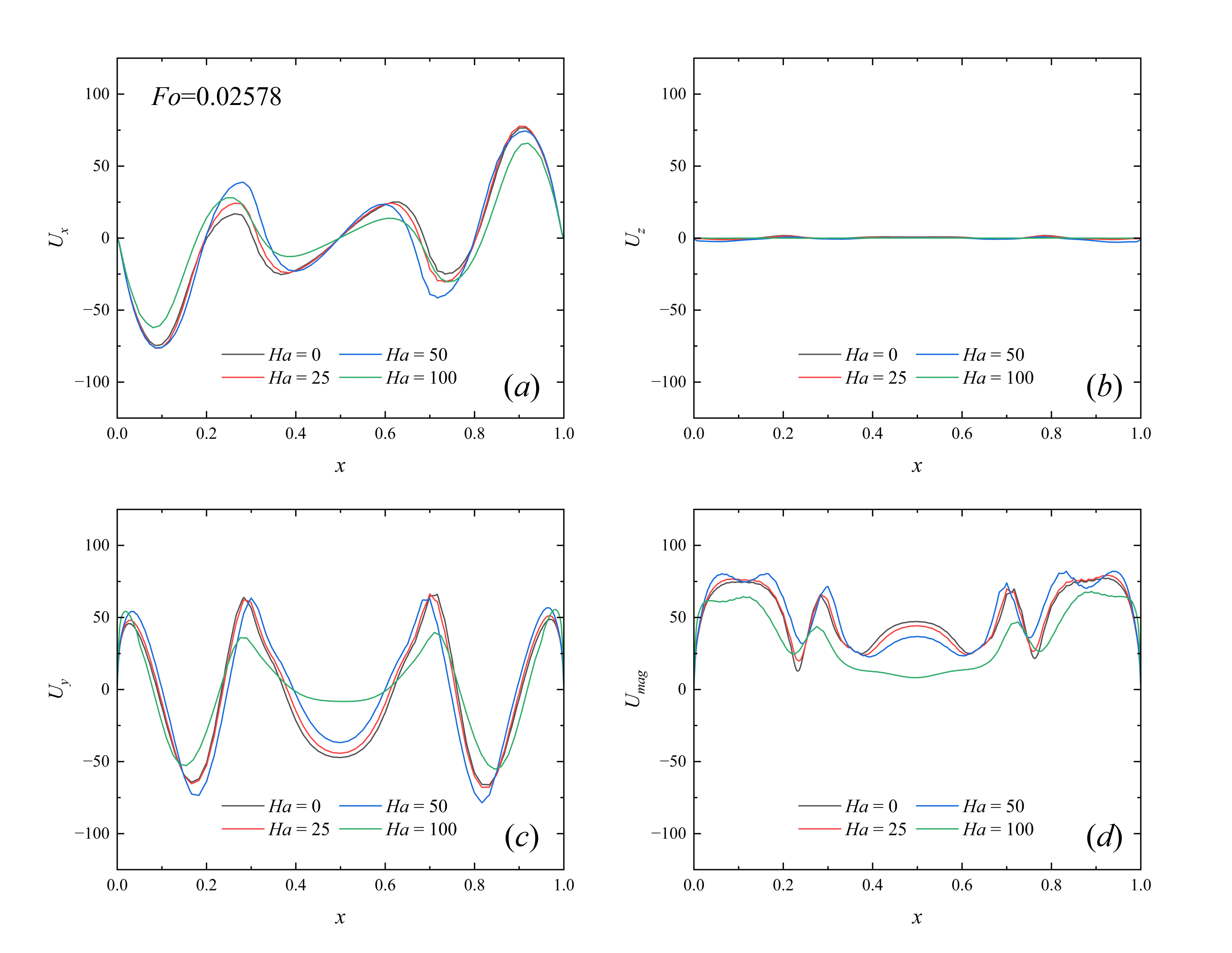}
 	\caption{\label{fig:fig9}Distributions of velocity components at the $z$-axis midline near the bottom heated wall ($y=0.05$, $z=0.5$) at $Fo=0.02578$ under different Hartmann numbers: (a) $U_x$ (b) $U_z$ (c) $U_y$ (d) $U_{mag}$.}
\end{figure}

To better illustrate the influence of magnetic field, dimensionless velocity profiles are plotted along the $x$-axis midline ($y=0.05$, $x=0.5$) near the heated wall and the $z$-axis midline ($y=0.05$, $z=0.5$) for analysis (Fig.~\ref{fig:fig8} and Fig.~\ref{fig:fig9}). As shown in Fig.~\ref{fig:fig8}, the velocity component $U_x$ remains close to zero for all $Ha$, while $U_y$ and $U_z$ exhibit oscillatory behavior along the $z$-axis. The fluctuations of $U_z$ are significantly larger near the sidewalls than at the center, indicating that the convection is primarily aligned with the $y\text{-}z$ plane and that horizontal flow is stronger near the walls.
As the Hartmann number increases, flow along the magnetic field direction is increasingly suppressed, leading to reduced convection intensity. Both $U_y$ and $U_z$ gradually decrease. In addition, the convection cells near the sidewalls are stretched along the magnetic field direction, while the central convection cells become compressed. Consequently, the number of convection cells decreases. At $Ha=0$ and 25, six convection cells are observed. At $Ha=50$, the number is reduced to two. When $Ha=100$, the convection cells aligned with the $y\text{-}z$ plane completely disappear, and Uz drops to nearly zero.

As shown in Fig.~\ref{fig:fig9}, the velocity component $U_z$ remains nearly zero, while $U_x$ and $U_y$ exhibit oscillations along the $x$-axis. This indicates that the convection at the $z$-axis midline ($y=0.05$, $z=0.5$) is primarily confined to the $x\text{-}y$ plane. For small Hartmann numbers, the value of $U_x$ between the cavity center and sidewall increases with increasing Hartmann number, while $U_x$ at the center and near the sidewalls remains nearly unchanged. In contrast, $U_y$ at the cavity center decreases as Hartmann number increases.
Combined with the streamline analysis, it can be inferred that the dominant circulation between the cavity center and sidewalls lies in the $x\text{-}y$ plane, while the central circulation is primarily in the $y\text{-}z$ plane. Under the influence of the magnetic field, convection along the $z$-direction is suppressed. As a result, the central circulation is weakened, while the near-wall circulation is enhanced. The magnetic field thus plays a rectifying role, which explains the observed velocity variations.
However, at $Ha=100$, this trend breaks down. At this point, the Lorentz force becomes dominant and suppresses the overall flow, leading to reduced velocities.

\subsubsection{\label{sec:level3}Effect of Magnetic Field on Flow and Heat Transfer Characteristics of the Melting Regimes}

\begin{figure}[]
	\centering
	\includegraphics[width=0.5\textwidth]{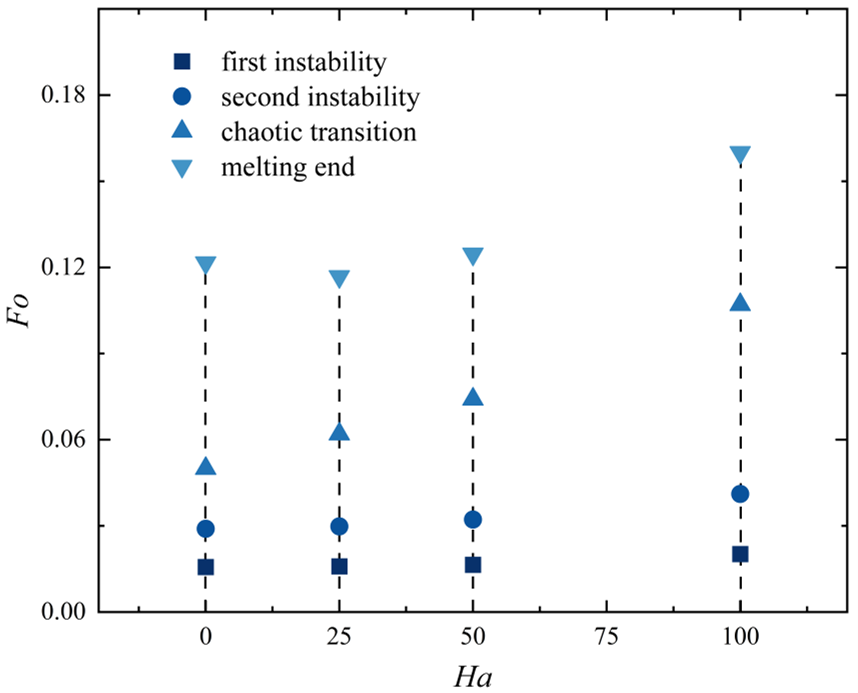}
	\caption{\label{fig:fig10}Critical dimensionless time for the onset of each melting regime under different Hartmann numbers.}
\end{figure}

\begin{figure}[]
 	\centering
 	\includegraphics[width=0.5\textwidth]{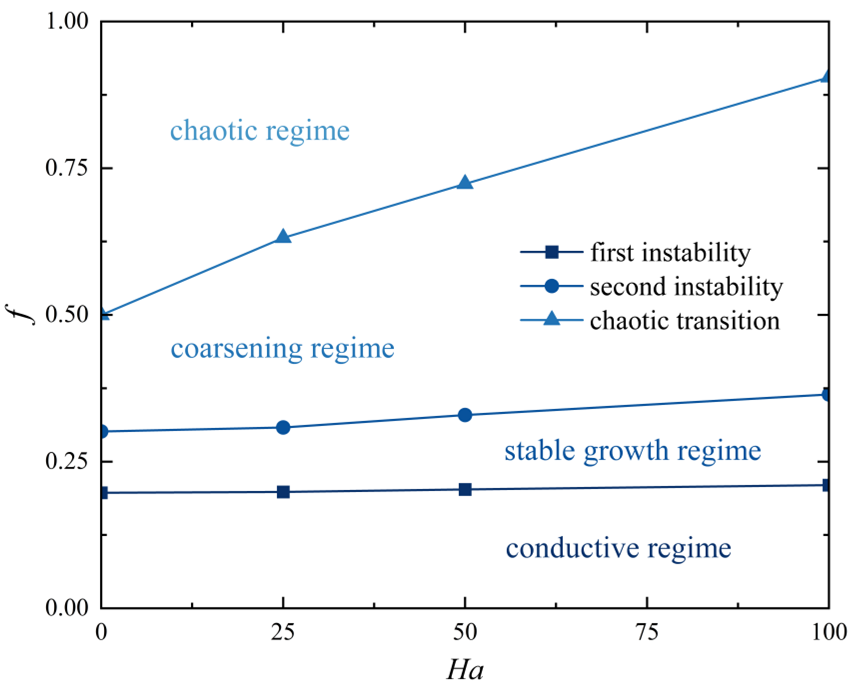}
 	\caption{\label{fig:fig11}Critical liquid volume fraction for the onset of each melting regime under different Hartmann numbers.}
 \end{figure}

Fig.~\ref{fig:fig10}
 and Fig.~\ref{fig:fig11}
  show the critical dimensionless time and corresponding liquid volume fraction marking the onset of each melting regime under different $Ha$. It can be observed that the magnetic field exerts a significant suppressing and stabilizing effect on the flow evolution. As the Hartmann number increases, the onset times of the convective, coarsening, and chaotic regimes are delayed, indicating that the system requires more time to transition into the next regime. This delay effect becomes more pronounced with increasing magnetic field strength, with the most substantial suppression observed in the coarsening regime. This is reflected by the increasing proportion of the coarsening regime in the overall melting process. It is worth noting that at $Ha$ = 25, the time required for complete melting is reduced (Fig.~\ref{fig:fig10}). This can be attributed to the fact that under a weak magnetic field, the Lorentz-force-induced flow rectification enhances convection in the main flow region, while the suppressive effect of the magnetic field on the flow remains relatively limited.

\begin{figure}[]
	\centering
	\includegraphics[width=0.7\textwidth]{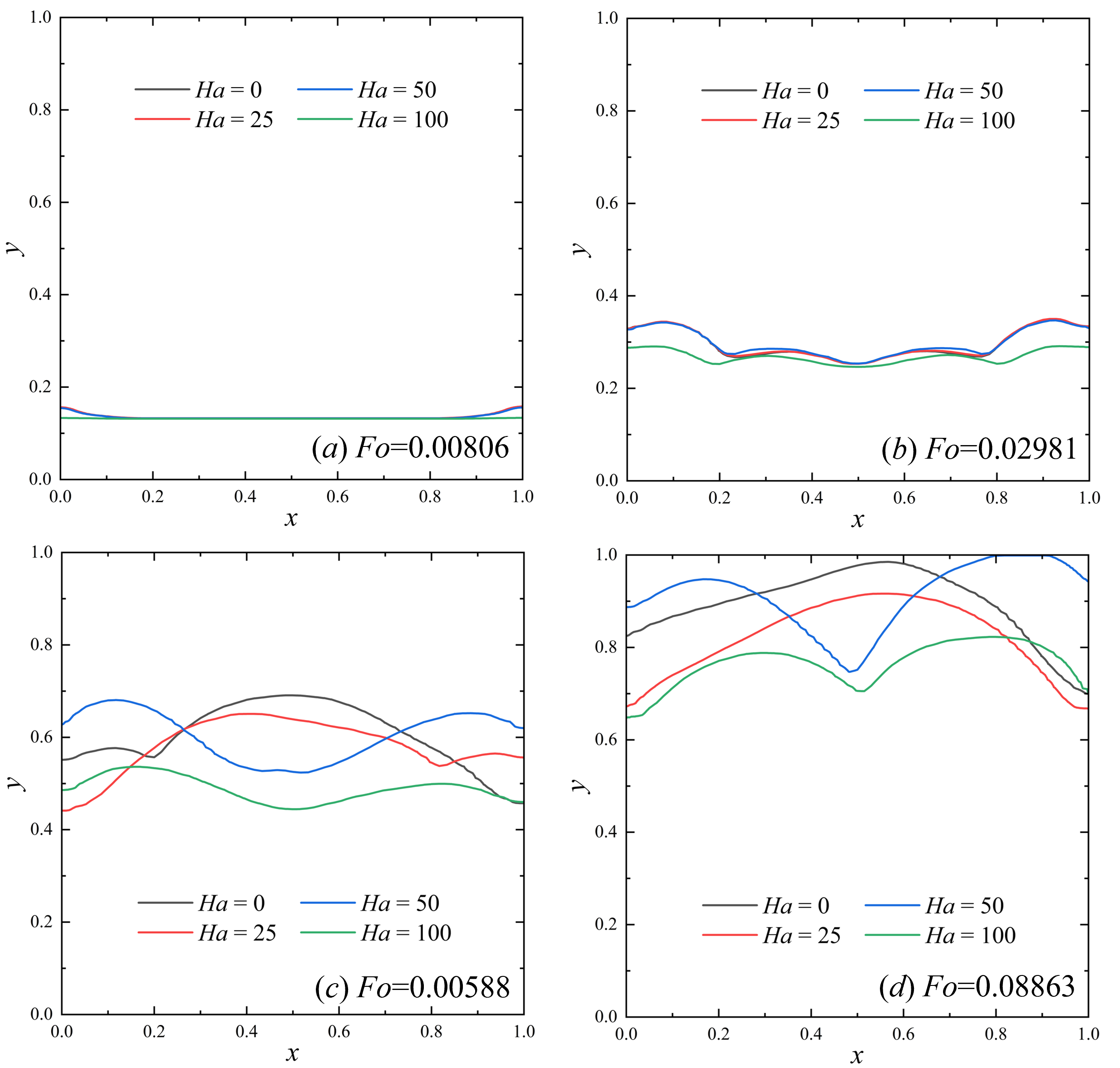}
	\caption{\label{fig:fig12}Positions of the solid-liquid interface in the $x\text{-}y$ plane ($z=0.5$) at $Fo$ of (a) 0.00806, (b) 0.02981, (c) 0.005882, and (d) 0.08863 under different Hartmann numbers.}
\end{figure}

\begin{figure}[]
	\centering
	\includegraphics[width=0.5\textwidth]{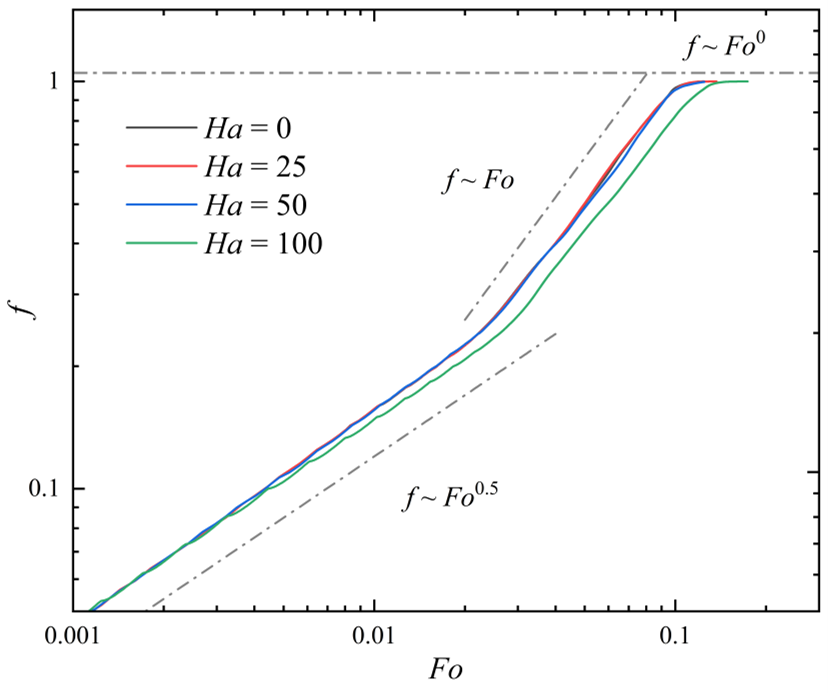}
	\caption{\label{fig:fig13}Evolution of the liquid volume fraction with dimensionless time. The gray line represents the scaling law between the liquid volume fraction and dimensionless time.}
\end{figure}

\begin{figure}[]
 	\centering
 	\includegraphics[width=0.5\textwidth]{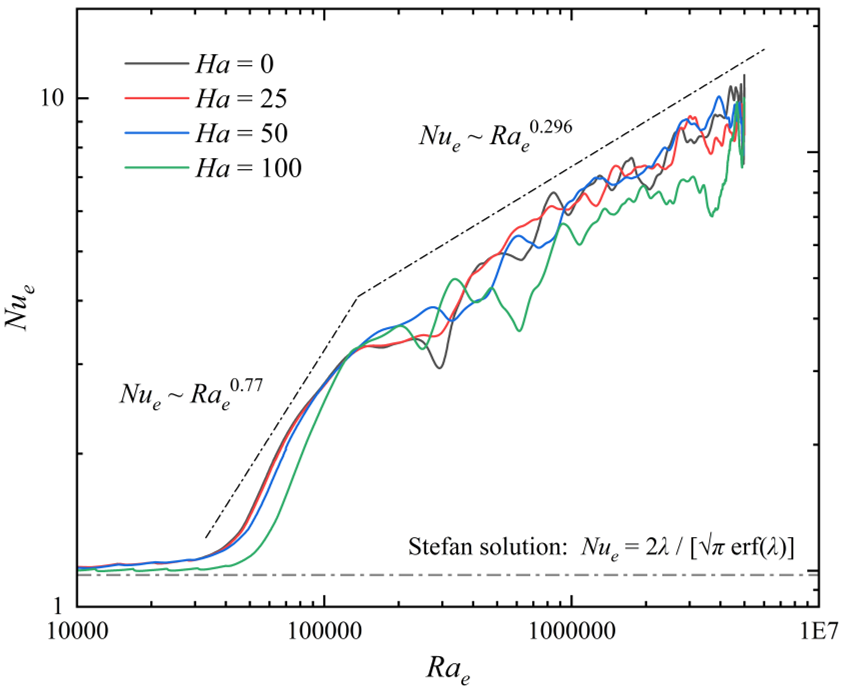}
 	\caption{\label{fig:fig14}Variation of the effective Nusselt number with the effective Rayleigh number under different Hartmann numbers.}
 \end{figure}

In the following, we present a detailed analysis of the magnetic field's impact on each melting regime.
During the conductive regime, the melt remains nearly stationary, the temperature is linearly distributed, and the solid–liquid interface is parallel to the bottom wall (Fig.~\ref{fig:fig12}
(a)). Fig.~\ref{fig:fig13}
 shows the evolution of the liquid volume fraction with dimensionless time under different magnetic field strengths. As shown, the melting rate in the conductive regime is not affected by the magnetic field strength, and all cases follow the scaling law $f \sim Fo^{0.5}$. In addition, the effective Nusselt number on the heated wall is only weakly influenced by the magnetic field during this regime and agrees with the Stefan solution (Fig.~\ref{fig:fig14}
 ).
As seen in Fig.~\ref{fig:fig10}, the influence of magnetic field strength in the conductive regime is relatively minor, resulting in only a slight delay in the onset of convection. The results show that the dimensionless time for the onset of convection is $Fo$=0.0157, 0.0159, 0.0164, and 0.0202 for $Ha$=0, 25, 50, and 100, respectively. The corresponding critical effective Rayleigh numbers are $Ra_c$=38222.929, 39032.375, 41603.238, and 46266.745, and the liquid volume fractions at the onset of convection are 19.699$\%$, 19.837$\%$, 20.264$\%$, and 20.994$\%$, respectively.
This trend can be explained by the fact that, in the presence of a magnetic field, the buoyancy force in the core region must overcome not only viscous and thermal diffusion effects but also the Lorentz force acting in the opposite direction of the flow. As the Hartmann number increases, the Lorentz force becomes stronger, requiring a higher effective Rayleigh number to initiate convection.

\begin{figure}[]
	\centering
	\includegraphics[width=0.7\textwidth]{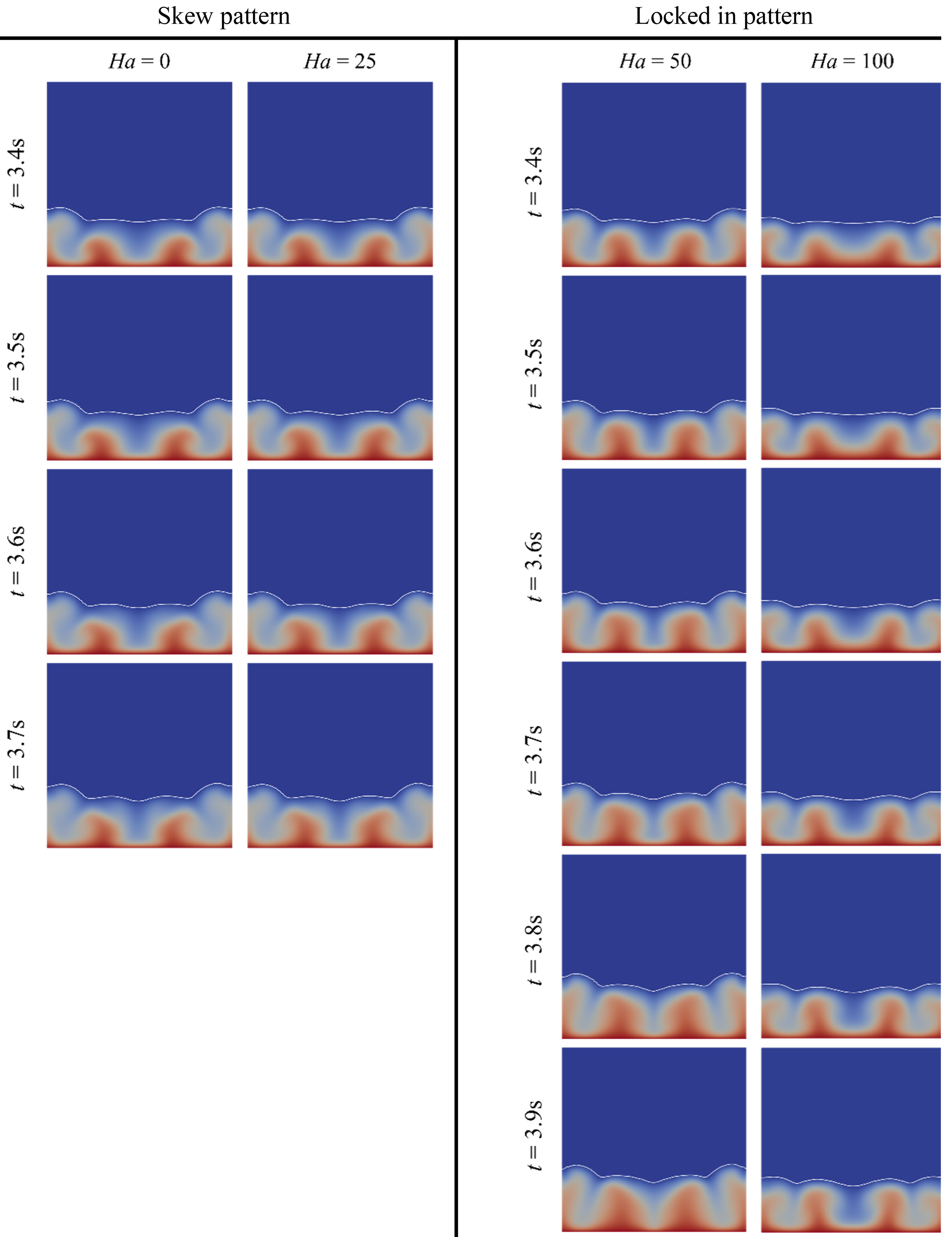}
	\caption{\label{fig:fig15}Temperature fields in the $x\text{-}y$ plane ($z=0.5$) at different times during the stable growth regime, showing two flow patterns under various Hartmann numbers.}
\end{figure}

\begin{figure}[]
	\centering
	\includegraphics[width=0.5\textwidth]{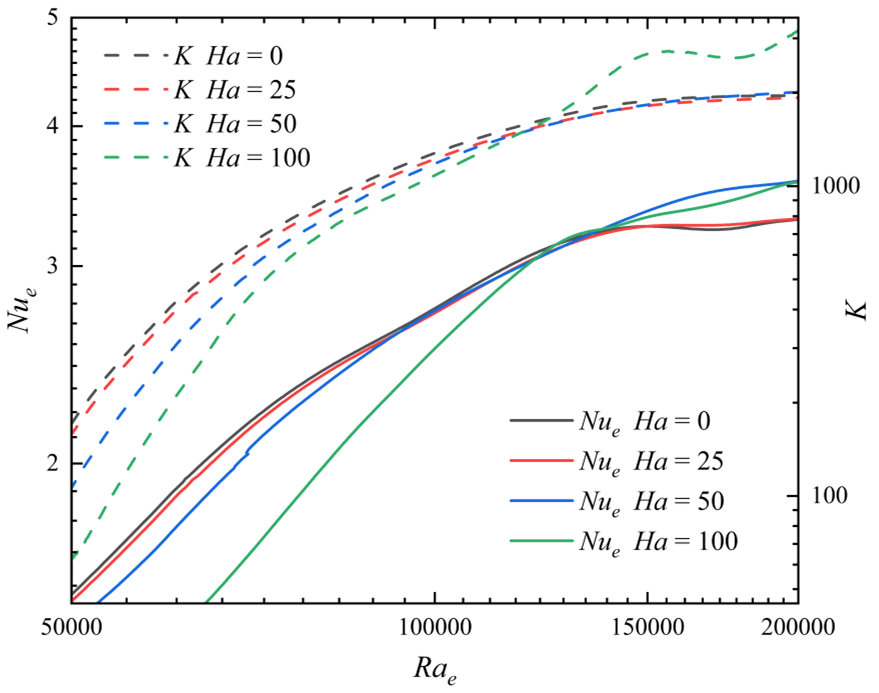}
	\caption{\label{fig:fig16}Variation of the effective Nusselt number (left axis) and average kinetic energy density (right axis) with the effective Rayleigh number during the stable growth regime under different magnetic field strengths. The black, red, blue, and green lines correspond to $Ha=0$, 25, 50, and 100, respectively.}
\end{figure}

During the stable growth regime, the influence of the magnetic field on flow and heat transfer becomes more significant. As convection develops, the scaling law between the liquid fraction f and the dimensionless time Fo changes to $f\sim Fo$, as shown in Fig.~\ref{fig:fig13}. In addition, the flow pattern also undergoes a transition. Fig.~\ref{fig:fig15}
 shows the flow patterns under different Hartmann numbers during this regime. In the middle and late stages of stable growth, we observe distinct flow patterns depending on the magnetic field strength.
When $Ha=25$, the flow pattern remains unchanged. The head of the thermal plumes bends horizontally, indicating a transition toward the coarsening regime. We refer to this as the skew pattern. However, when $Ha=50$ and 100, a new flow pattern emerges: the convective cells become locked beneath the wavy solid-liquid interface and grow vertically in a stable manner over time, without lateral deflection. This flow pattern has also been observed in RB phase change systems without magnetic fields,\cite{lu_rayleigh-benard_2021,vasil_dynamic_2011} and is referred to as the locked in pattern.
Lu et al.\cite{lu_rayleigh-benard_2021} investigated RB phase change systems without magnetic fields and found that the locked-in pattern is caused by the non-planar topography of the upper melting boundary. However, when the timescale separation is large, the convective cells enter an unstable nonlinear equilibrium. In this case, the locking effect of the topography weakens, and the flow becomes more unstable. This equilibrium is easily disrupted, as the fluid cannot sustain continuous vertical stretching. As a result, the flow may quickly bifurcate into a new, more unstable set of convective cells, even if the mean fluid depth has not changed significantly. This breakdown of the locked in pattern leads to noticeable deformation of the interface topography.
The previous analysis indicates that the magnetic field suppresses convection in the direction parallel to the field, leading to a transition of the flow pattern from a 3-D convective cell pattern to a quasi-2-D convective roll pattern with the roll axes aligned parallel to the magnetic field. As the magnetic field strength increases, this suppressive effect becomes stronger, causing the evolution of convection to occur on a slower timescale and resulting in a reduced timescale separation. Consequently, under high Hartmann numbers, the convective cells become locked beneath the wavy solid-liquid interface and grow vertically, giving rise to the locked in pattern.
As shown in Fig.~\ref{fig:fig13}, the melting rate increases after the onset of convection, and is only slightly affected by the magnetic field. Therefore, the position of the solid–liquid interface in this regime is mainly determined by the timing of convection onset. From Fig.~\ref{fig:fig12}(b), it can be seen that the interfaces for $Ha=0$ to 50 nearly overlap, while for $Ha=100$, the interface develops more slowly due to the delayed onset of convection. The shape of the solid-liquid interface is mainly governed by the distribution of thermal plumes. The interfacial wavenumber equals the number of thermal plumes, which remains four during the stable growth regime.
Fig.~\ref{fig:fig16}
 shows the variation of the effective Nusselt number $Nu_e$ and average kinetic energy density $K$ with the effective Rayleigh number $Ra_e$ under different magnetic field strengths during the stable growth regime. In the early stage of this regime, magnetic suppression dominates, and the average kinetic energy density decreases with increasing magnetic field strength. In the later stage, under the locked-in pattern, the central convective rolls do not exhibit horizontal deflection, reducing horizontal energy dissipation and enhancing the main circulation in the core region. As a result, the average kinetic energy density for $Ha=50$ and 100 increases. Moreover, the locked-in pattern extends the vertical extent of the central convective region, promoting direct heat transfer between the bottom and top boundaries. This leads to thinner thermal boundary layers and a higher effective Nusselt number at the heated wall.

During the coarsening regime, the development and distribution of convective cells and thermal plumes are strongly affected by the magnetic field strength. When $Ha=0$ and 25, it is observed that in the early stage of coarsening, the two central plumes expand while the plumes near the sidewalls shrink. Subsequently, the central two plumes begin to merge, and the sidewall plumes grow larger. Eventually, the central plumes fully merge into a single large plume, accompanied by two smaller side plumes, resulting in a total of three plumes. During this process, both $K$ and $Nu_e$ remain nearly constant.
In the late stage of coarsening, the central large plume and one of the side plumes expand upward, while the other side plume decays. This leads to a final state consisting of one large central plume and one small side plume. Both $K$ and $Nu_e$ exhibit step-like increases during this stage.
When $Ha=50$, the central plumes first merge, and the side plumes in the direction perpendicular to the magnetic field decay, while those parallel to the field grow. Afterwards, the central plume shrinks, and the side plumes continue to expand. Eventually, thermal plumes exist only near the sidewalls parallel to the magnetic field, and the flow transitions to a quasi-2-D convective roll pattern.
For $Ha=100$, only one coarsening event occurs, where the central plume merges with the side plume in the direction of the magnetic field. The final state features two columns of thermal plumes near the sidewalls aligned with the field, and the flow exhibits a fully quasi-2-D convective roll pattern. As melting proceeds, these rolls grow upward, and the majority of melting takes place during this regime (see Fig.~\ref{fig:fig11}). The number and distribution of plumes after coarsening can be inferred from the topography of the solid–liquid interface (Fig.~\ref{fig:fig12}(c)).
The study reveals that during the chaotic stage, thermal plumes shift and then collapse, while irregular thermal plumes unrelated to convection emerge near the walls. As shown in Fig.~\ref{fig:fig14}, $Nu_e$ exhibits oscillatory growth with increasing $Ra_e$ in the chaotic stage. 

To further analyze the effect of the magnetic field on $Nu_e$, a linear regression was performed on the logarithm of the results. The fitting results indicate that after the onset of convection, the influence of a small magnetic field ($Ha\le50$) on the effective Nusselt number is minor and can be neglected. Conversely, under a large magnetic field ($Ha=100$), the flow and heat transfer are significantly suppressed by the magnetic field. This conclusion is consistent with the preceding analysis of velocity (Fig.~\ref{fig:fig8} and ~\ref{fig:fig9}).
In the early to middle stages of the stable growth phase, as convection initially develops, the effective Nusselt number increases rapidly (Fig.~\ref{fig:fig14}). For $Ha\le50$, the fitted correlation is $Nu_e=0.00037Ra_{e}^{0.7703} (1+Ha)^{-0.0058}$. For $Ha=100$, the fitted correlation is $Nu_e=\num{3.6e-5}Ra_{e}^{0.9659}$.
In the late stable growth phase, the coarsening phase, and the chaotic phase, convection is fully developed, and the growth rate of the effective Nusselt number decreases (Fig.~\ref{fig:fig14}). For $Ha\le50$, the fit yields the correlation $Nu_e=0.0993Ra_{e}^{0.296} (1+Ha)^{-0.0027}$. For $Ha=100$, where flow and heat transfer are significantly suppressed, the fitted correlation is $Nu_e=0.0705Ra_{e}^{0.3107}$.

\section{\label{sec:level1}conclusion}

In this study, the melting of solid gallium in a bottom-heated and top-cooled cubic cavity under a horizontal magnetic field was investigated through numerical simulations. The dynamical behavior of solid–liquid phase change in a low-Prandtl-number Rayleigh-Bénard system was systematically analyzed for different Hartmann numbers ($Ha = 0$, 25, 50, and 100). The primary findings are summarized below.

In the absence of a magnetic field, the melting process can be clearly divided into four stages: the conduction stage, the stable growth stage, the coarsening stage, and the chaotic stage. During the conduction stage, the system is dominated by thermal conduction. The solid-liquid interface remains flat, and no significant convection is observed. When the effective Rayleigh number Rae reaches the critical value, buoyancy-driven convection begins to develop, marking the onset of the stable growth stage. Uniformly distributed thermal plumes appear, and the interface becomes periodically wavy due to accelerated local melting. As melting proceeds, adjacent thermal plumes merge, reducing the number of convective cells and transitioning the system into the coarsening stage. This stage is characterized by strong fluctuations in Nusselt number Nue and kinetic energy, with no clear trend. Eventually, the flow evolves into an irregular chaotic state, accompanied by severe deformation of the interface. In this stage, the effective Nusselt number follows the scaling law $Nu_e\sim Ra_{e}^{0.29}$.

When a horizontal magnetic field is applied, the flow, heat transfer, and melting behavior undergo significant changes:
\begin{itemize}
	\item Quasi-2-D flow transition: The Lorentz force suppresses flow in the direction parallel to the magnetic field, driving a transition from the 3-D convective cell pattern to a quasi-2-D convective roll pattern. As the Hartmann number increases, this transition becomes more pronounced. At $Ha=100$, the flow fully transforms into rolls aligned with the magnetic field.
	\item Suppression and delay effects: The magnetic field has a strong stabilizing effect on flow development, delaying the onset of the convective, coarsening, and chaotic stages. A higher Hartmann number corresponds to a larger critical effective Rayleigh number required to trigger flow instability.
	\item Emergence of the locked in pattern: During the stable growth stage at high Hartmann numbers ($Ha=50$ and 100), a "locked in pattern" is observed. Convective cells are confined beneath the wavy interface and grow vertically without horizontal bending. This pattern enhances vertical heat transfer in the central region, resulting in a thinner thermal boundary layer and a higher effective Nusselt number.
\end{itemize}

In summary, this study provides an in-depth investigation into the flow and heat transfer characteristics of a three-dimensional phase change Rayleigh-Bénard system under low Prandtl number conditions and a horizontal magnetic field. Scaling relations were established to quantify the melting dynamics and heat transfer performance, offering potential applications in thermal energy storage, electronic cooling, and electromagnetic metallurgy. While the present work focuses on horizontal magnetic fields, future research could explore the effects of vertical or inclined magnetic fields on the system’s dynamics. Particular attention should be given to how different magnetic field configurations influence flow pattern, interface morphology, and heat transfer efficiency.

\begin{acknowledgments}
	The work was supported by the National Science Foundation of China (52376155) and the Fundamental Research Funds for the Central Universities.
\end{acknowledgments}

\section*{Conflict of Interest}

The authors have no conflicts to disclose.

\section*{Author Contributions}

Xinyi Jiang: Conceptualization (equal); Data curation (equal); Investigation (equal); Methodology (equal); Writing –
original draft (equal); Writing – review \& editing (equal).
Chenyu You: Validation (equal); Writing – review \& editing (supporting).
Xinning Nan: Writing – review \& editing (supporting).
Jiawei Chang: Writing – review \& editing (supporting).
Zenghui Wang: Conceptualization (equal); Funding acquisition (equal); Project administration (equal); Resources (equal); Supervision (equal); Writing – review \& editing (equal). 

\section*{DATA AVAILABILITY}

The data that support the findings of this study are available from the corresponding author upon reasonable request.


%
%

%


\bibliography{references}

\end{document}